%% file: main.tex
\documentclass[manuscript]{acmart-arxiv}
\usepackage{booktabs} 
\usepackage{algorithm, algorithmic}
\usepackage{caption}
\usepackage{subcaption}
\usepackage{cleveref}
\usepackage{mathtools}
\DeclareMathOperator{\EX}{\mathbb{E}}

\begin{document}




\title{Exploring Low-degree Nodes First Accelerates Network Exploration}


\author{Stefania Costantini}
\orcid{0000-0002-5686-6124}
\email{stefania.costantini@univaq.it}
\affiliation{%
  \institution{University of L'Aquila}
  \department{Department of Information Engineering, Computer Science and Mathematics}
  \city{L'Aquila}
  \postcode{67100}
  \country{Italy}
}
\author{Pasquale De Meo}
\authornote{Corresponding author}
\orcid{0000-0001-7421-216X}
\email{pdemeo@unime.it}
\affiliation{%
  \institution{University of Messina}
  \department{Department of Ancient and Modern Civilizations}
  \city{Messina}
  \postcode{98166}
  \country{Italy}
}

\author{Angelo Giorgianni}
\author{Valentina Migliorato}
\affiliation{%
  \institution{University of Messina}
  \department{Department of Ancient and Modern Civilizations}
  \city{Messina}
  \postcode{98166}
  \country{Italy}
}

\author{Alessandro Provetti}
\orcid{0000-0001-9542-4110}
\email{ale@dcs.bbk.ac.uk}
\affiliation{%
  \institution{Birbeck, University of London}
  \department{Department of Computer Science and Information Systems}
  \city{London}
  \postcode{WC1E 7HX}
  \country{UK}
}

\author{Federico Salvia}
\affiliation{%
  \institution{University of Messina}
  \department{Department of Ancient and Modern Civilizations}
  \city{Messina}
  \postcode{I-98166}
  \country{Italy}
}

\renewcommand{\shortauthors}{Costantini et al.}

\begin{abstract}
We consider information diffusion on Web-like networks and how random walks can simulate it. 
A well-studied problem in this domain is Partial Cover Time, i.e., the calculation of the expected number of steps a random walker needs to visit a given fraction of the nodes of the network.
We notice that some of the fastest solutions in fact require that nodes have perfect knowledge of the degree distribution of their neighbors, which in many practical cases is not obtainable, e.g., for privacy reasons.  
We thus introduce a version of the Cover problem that considers such limitations: Partial Cover Time with Budget. 
The budget is a limit on the number of neighbors that can be inspected for their degree; we have adapted optimal random walks strategies from the literature to operate under such budget. 
Our solution is called Min-degree (MD) and, essentially, it biases random walkers towards visiting peripheral areas of the network first. 
Extensive benchmarking on six real datasets proves that the---perhaps counter-intuitive strategy---MD strategy is in fact highly competitive wrt. state-of-the-art algorithms for cover.
\end{abstract}

%
%

\maketitle

\input{introduction}
\input{background}

\input{related-works}
\input{approach_description}

\input{experiments}

\input{conclusions}

\begin{acks}
This article is based upon work from COST Action DigForAsp CA17124, supported by COST (European Cooperation in Science and Technology. \url{www.cost.eu}
\end{acks}

\bibliographystyle{ACM-Reference-Format}
\bibliography{bibliography}
\end{document}

%% file: introduction.tex
\section{Introduction}
\label{sec:intro}

A number of data sources on the Web can be described as {\em network data,} i.e., collections of interrelated, often heterogeneous, objects (people, documents, multimedia objects and so on) tied by some kind of relationships. 
Important examples of network data are the friendship network in Facebook \cite{viswanath2009evolution}, mutual following relationships in a software development platform as GitHub \cite{rozemberczki2019multi} or co-purchase relationships between members of an e-commerce Web site like Amazon \cite{yang2015defining}.
In what follows we will speak interchangeably of networks and graphs \cite{newman2018networks} as an ordered pair $G = \langle N,E\rangle$ consisting of a collection of nodes (associated to artificial or real entities) and edges (which capture relationships between nodes).

Random Walks \cite{lovasz1993random} are an important class of algorithms to analyze the structure of large networks; in short, a random walk on a graph can be described as a random process which starts from one of the graph nodes and, in a sequential fashion, selects the next node to move according to some specified probability \cite{lovasz1993random}. 
Random walks (RWs) have been applied in a broad range of graph analytic tasks such as the ranking of individuals in a social network \cite{newman2005measure} and the segmentation of large virtual communities \cite{PonsL06,MeoFFP14}. 
One of the most important application of RWs is {\em network sampling} \cite{hu2013survey}: a family of techniques that takes a graph $\mathcal{G}$ and seek to generate a representative subgraph $\mathcal{G}^{\prime}$ which preserves some of the structural properties of $G$. Graph sampling has a wide spectrum of applications on the Web such as the identification of a sample of people to poll from an hidden population in sociological studies \cite{hu2013survey}, or the crawling of large Online Social Networks \cite{Ahn*07,Catanese*11,gjoka2011practical}.

Many studies focused on estimating the {\em efficiency} of random walks and several parameters have been introduced so far \cite{Aleliunas*79,kahn1989cover,redner2001guide, avin2008power,ikeda2009hitting}.
A key parameter to assess the efficiency of a random walk is {\em partial cover time} \cite{avin2004efficient,avin2005cover,weng2017partial,chupeau2015cover}, which quantifies the time a RW takes to visit a  given fraction of the nodes of $G$. 
Currently, the main focus in literature has been the ``extremal'' version of the problem, {\em cover time} \cite{aldous1989introduction} defined as the expected number of steps a RW needs to visit all nodes in $G$. 
Fewer studies have addressed cover time, mostly focusing on the boundary cover time for specific classes (e.g. regular graphs) \cite{kahn1989cover} or on heuristics \cite{abdullah2015speeding,ikeda2009hitting}.

We submit that in Web-based applications the (total) cover time may not be an interesting indicator vis-\`a-vis an optimized (or at least reduced) partial cover time. 
For instance, consider rumor spreading  in Online Social Network: we are not worried if the rumor reached the entire population but we strive to spread the rumor to a sufficiently large sample of the whole population. 

Of course, existing solutions for cover time might be extended to the partial cover time but they make assumptions which we believe are unrealistic in Web applications. 
For instance, the approach of \cite{ikeda2009hitting} requires that a node knows the degrees of all of its neighbors to compute the probability that a random walk move from a node to one of its neighbours. 
In Web-based applications such as Online Social Networks, an individual may refuse to disclose the number of and identities of her/his friends for privacy reasons.

In this paper we introduce a new problem, called {\em Partial Cover Time with Budget} in which we wish to design a random walk whose partial cover time is as small as possible under the constraint that any node in the graph is allowed to query only a random sample of fixed size of its neighbors to retrieve their degrees.

We propose a new algorithm for reducing the partial cover time, called {\em Min-Degree} (in short, \texttt{MD}). 
The \texttt{MD} algorithm combines ideas from the literature in a novel way, which builds random walks  displaying these two key properties:  {\em (a)} RWs will preferentially visit unvisited nodes first and {\em (b)}  among unvisited nodes, RWs will prefer transitioning to the lowest-degree node.

We have conducted extensive validation tests on  six real-life graph.
On each we have compared the \texttt{MD} algorithm with four state-of-the-art algorithms and found it highly competitive.

This paper is organized as follows. 
In Section \ref{sec:background} we provide basic definitions and background results while in Section \ref{sec:related-works} we discuss the related literature.
Section \ref{sec:approach-description} describes the \texttt{MD} algorithm while we present the main findings of our experimental analysis in Section \ref{sec:experiments}. 
Finally, we draw our conclusions in Section \ref{sec:conclusions}.

%% file: background.tex
\section{Background}
\label{sec:background}

Let $\mathcal{G}= \langle N ,E \rangle$ be an undirected and connected graph with $\vert N \vert = n$ nodes and $\vert m \vert = m$  edges.
We say that $G$ is of {\em order} $n$ and {\em size} $m$.

For any node $i \in N$, let $d_i$ be the degree of $i$, i.e., the number of edges incident onto $i$ and let $N(i)$ be the set of neighbours of $i$, i.e, the set of nodes $j \in N$ for which the edge $\langle i, j \rangle$ belongs to $E$.

A Random Walk (in short, RW) on the graph $G$ is the process of visiting the nodes of $G$ in some sequential random order. The RW starts at some fixed node, and, at each step, it moves from the current node (say $i$) to the next one (say $j$) with probability (called {\em transition probability}) $p_{ij}$.  We can collect $p_{ij}$ transition probabilities into a matrix $\mathbf{P}$ called {\em transition matrix probability}.

A {\em Simple Random Walk - SRW} is 
a Random Walk such that the next node to visit is chosen uniformly at random from the set of neighbors of the current node. In other words, if the walk is at node $i$, then it will move to the node $j$ in the next step with probability $p_{ij} = \frac{1}{d_i}$ if $j \in N(i)$ and $p_{ij} = 0$ otherwise.

Let us consider a RW starting from a node, say $i$: we say that the RW {\em covers} $G$ if the RW visits at least once every node in $G$ \cite{kahn1989cover}.
For each node $i \in N$ we can define a random variable $X_i$ which specifies the first time a RW starting from $i$ covers $G$.

A long standing problem in random walk theory consists of estimating the expected value of $X_i$, for any node $i$ from which the random walk starts visiting $G$.

More formally, we provide the following definition \cite{aldous1989introduction,kahn1989cover}:

\begin{definition}[Cover Time]
Given a node $i$ , the {\em cover time} $C_G(i)$ for the node $i$ is defined as $C_G(i) = \EX[X_i]$, i.e., it is the {\em expected number of steps} the random walk takes to visits all nodes in $\mathcal{G}$, provided that it starts from $i$. 

The {\em maximum cover time} $C_G$ is defined as:

\begin{equation}\label{eqn:cover}
C_G = \max_{i \in N} C_G(i)
\end{equation}

\end{definition}

The cover time of a graph represents thus a parameter to evaluate the efficiency of a random walk, i.e., to quantify how fast a random walk is in covering $\mathcal{G}$.
The cover time of a graph (along with methods for bounding it) have been extensively investigated \cite{Aleliunas*79,kahn1989cover,matthews1988covering, chandra1996electrical}, especially for Simple Random Walks. 

One of the first result is due to Aleliunas {\em et al.} \cite{Aleliunas*79} who showed that for any connected graph $G$ the cover time $C(G)$ satisfies $C(G) < 2\times m \times n$ which is bounded above by $O(n^3)$. 
Feige \cite{feige1995tightlower,feige1995tightupper} improved the results of \cite{Aleliunas*79} and, specifically, he showed that, for any connected graph $G$, the cover time satisfied the following condition:

\begin{equation}
\label{eqn:bounds}
\left(1 - o(1)\right) n \log n < C_G < \left(1 + o(1)\right)\frac{4}{27}n^3
\end{equation}

The lower bound occurs in case of a complete graph of order $n$ (i.e. a graph in which any pair of nodes is connected by an edge) while the upper bound occurs for the so-called {\em lollipop graph}.
In case of regular graphs (i.e., graphs in which nodes have the same degree), Kahn {\em et al.} \cite{kahn1989cover} proved that $C(G)$ is bounded above by $O(n^2)$. 

In general, highly connected graphs display the lowest cover time; in contrast, if graph connectivity is poor or if bottlenecks exist in the graph, then we expect an increase in cover time.

In many Web-based applications, however, the cover time may not be a reliable indicator of the efficiency of a random walk. 
For instance, suppose we consider a virtual community and let us focus on the spreading of a rumor in that community; in general, it does not matter that low-degree nodes receive the rumor and it does not matter that the whole population receives the rumor. 
In many cases, it suffices to verify that a relatively large portion of the whole population has received that rumor and, thus, we are required to estimate the number of steps a walk takes before visiting a fraction $\tau$ (with $0  \leq \tau \leq 1$) of nodes in $\mathcal{G}$. 
Such an intuition is encoded in the notion of {\em partial cover time} \cite{avin2004efficient, avin2008power}:

\begin{definition}[Partial Cover Time]
\label{def:partial-cover-time}
Let $\mathcal{G}$ be undirected and connected with order $n$ and  let $i$ be a node in $G$ and $\tau \in [0, 1]$. 
The {\em partial cover time} $PCT_G(\tau, i)$ for node $i \in N$ is the expected number of steps a random walk takes to visit at least $\lfloor \tau  \times \vert N \vert \rfloor$ nodes in $\mathcal{G}$, provided that the random walk starts from the node $i$.
The {\em partial cover time} $PCT_G(\tau)$ is defined as follows.

\begin{equation}
\label{eqn:partial-cover-time}
PCT_G(\tau) = \max_{i \in N} PCT_G(\tau, i)
\end{equation}
\end{definition}

Some important bounds on $PCT_G(\tau)$ are possible, as in the the following.

\begin{definition}[Hitting Time]
Let $\mathcal{G} = \langle N, E \rangle$ be an undirected and connected graph.
Given a pair of nodes $i \in N$ and $j \in N$, the {\em hitting time} $H_G(i, j)$ is defined as the {\em expected number of step} a random walk takes to get to $j$, provided that it starts from $i$. 
The {\em maximum hitting time} $H_G$ is defined as:

\begin{equation}
\label{eqn:hitting}
H_G = \max_{i \in N, j \in N} H_G(i,j)
\end{equation}
\end{definition}

\cite{avin2004efficient} proved that for any graph $\mathcal{G}$ and $0 \leq \tau < 1$ we have that $PCT_G(\tau) \in \Theta(H_G)$.
As a consequence, if $\mathcal{G}$ is such that $H_G \in O(n)$, then there exists a random walk which achieves a partial cover time which is also linear in the number $n$ of graph nodes.

\cite{avin2004efficient} considered a partial cover time in the order of $O(n)$ as {\em optimal} and they provided some examples of graphs for which it is possible to design random walks achieving optimal partial cover time, namely i) the complete graph, ii) the star, iii) the hypercube, iv) the 3-dimensional mesh and v) random geometric graphs (i.e., undirected graphs where nodes belong to some metric space and the probability of an edge between two nodes decreases with their distance in that space).

%% file: related-works.tex
\section{Related Works}
\label{sec:related-works}

In this section we review some of the most popular techniques to reduce the cover time of a random walk.

\subsection{Non-uniform transition probabilities}
\label{sub:non-uniform}

Some authors \cite{ikeda2009hitting,abdullah2015speeding} suggested to use proper transition probabilities, which derive from the knowledge of the topology of $\mathcal{G}$, to reduce the cover time $C(\mathcal{G})$.

In detail, a very important result is due to Ikeda {\em et al.} \cite{ikeda2009hitting}, who considered a transition probability matrix $\mathbf{P}$ defined as follows:

\begin{equation}
\label{equ:transition-iked}
p_{ij}=
\begin{cases}
 \frac{d_j^{-1/2}}{\sum_{k \in N(i)} d_j^{-1/2}}, & \text{if } \langle i, j \rangle \in E\\
 0, & \text{otherwise } 
\end{cases}
\end{equation}

\cite{ikeda2009hitting} proved that, {\em for any graph} $G$, a random walk in which transition probabilities follow Equation \ref{equ:transition-iked} has an hitting time in the order of $O(n^2)$ and a cover time in the order of $O(n^2 \log n)$. 
\cite{ikeda2009hitting} proved also that a random walk whose transition probabilities obeyed Equation \ref{equ:transition-iked} were also {\em optimal} for graphs with an {\em arbitrary topology}, i.e., it is not possible to further reduce the cover time unless we restrict our attention on special classes of graphs.
The results of Ikeda {\em et al.} \cite{ikeda2009hitting} assumes that each node knows the degree of all its neighbors. In addition, observe that random walk in the framework of \cite{ikeda2009hitting} are no longer simple because the walker may cross a node more than once; intuitively, such an approach works because a node tend to privilege low degree neighbours, thus favouring the exploration of regions of $G$ which would be hard to reach.

Abdullah {\em et al.} \cite{abdullah2015speeding} suggested as to use transition probabilities of the form $p_{ij} \propto 1 / \min\left(d_i, d_j\right)$ and they called their choice {\em the minimum degree weighting scheme}. For this choice of transition probabilities, Abdullah {\em et al.} \cite{abdullah2015speeding} proved that for every connected graph the hitting time is at most $6n^2$ that the cover time is at most $O(n^2 \log n)$. They further conjectured that if the minimum degree weighting scheme is applied, then every connected graph has cover time $O(n^2)$ but such a conjecture is still unverified to our knowledge.


\subsection{Random Walks which prefer unvisited edges}

An important research avenue to reduce cover time is to consider modified random walks which record the edges the random walk used to explore the graph $G$. More specifically, suppose that a particular step the random walk occupies a node $i$ and let us consider the set of edges incident onto $i$.
If there is at least an {\em unvisited edge} (i.e. an edge which has never been used by the random walk to explore $\mathcal{G}$), then the random walk picks one of the unvisited edges according to a prescribed rule $\mathcal{A}$; if there are no unvisited edges incident onto  the  node currently occupied by the random walk, then the random walk moves to a random neighbour.

The process above is called {\em E-Process} (or edge-process) \cite{BeCoFr15}. In the simplest case, the rule $\mathcal{A}$ is a uniform random choice over unvisited edges incident onto the node currently occupied by the walker but we do not exclude arbitrary choices of $\mathcal{A}$; as highlighted in \cite{BeCoFr15}, the rule could be determined on-line by an adversary, or could vary from node to node.

An important approach to cite is due to Avin and Krishnamachari \cite{avin2008power}, who
explicitly focused on the reduction of the partial cover time. \cite{avin2008power}
introduced the so-called {\em Random Walk with Choice}, or in short, $RWC(d)$ algorithm. 
The $RWC(d)$ algorithm is an extension of a standard random walk and, specifically, if we suppose that the random walk reaches a node $i$ at the time step $t$, then the $RCW(d)$ algorithms performs the following steps:

\begin{enumerate}
\item It selects, uniformly at random and with replacement, $d$ of the neighbors of $i$, say $D(i)$ with $\vert D(i) \vert = d$.
\item The random walk moves to the node $j$, selected according to the following rule:
\begin{equation}
\label{equ:rwc-selection}
j = \arg \min_{j \in D(i)} \frac{c_t(j) + 1}{d_j} 
\end{equation}
\end{enumerate}
Here $c_t(j)$ counts the number of times the node $j$ has been visited up to the time step $t$.

The parameter $d$ is determined through experiments but in the special case $d = 1$ the $RWC(d)$ algorithm coincides with a Standard Random Walk.

%% file: approach_description.tex
\section{Approach Description}
\label{sec:approach-description}

We now present our approach, called {\em Min-Degree} (or, in short, \texttt{MD}) to reduce the partial cover time of an undirected and connected graph $G = \langle N, E \rangle$.

\subsection{Main Features of the \texttt{MD} algoritmh}
\label{sub:main-features}

Previous research findings are relevant to design efficient strategies to navigate $\mathcal{G}$, i.e., strategies that use the lowest number of steps to visit a fraction $\tau$ of the nodes.
For instance, the procedure proposed by \cite{ikeda2009hitting} is {\em optimal} for the cover time, in the sense that if we would choose transition probabilities as in Equation \ref{equ:transition-iked} then we would obtain a random walk whose cover time is $O(n^2 \log n)$: the best lower bound for cover time we can hope for. 

Unfortunately, the approach of \cite{ikeda2009hitting} requires that a node knows the degrees of all of its neighbors. 
In the Social Web scenario (and, in general, in many Web related domains) such an assumption may be unrealistic: for instance, in real Online Social Networks, an individual may refuse to disclose the number and the identities of her/his friends for privacy reasons; in addition, for some applications, the time required for generating the full list of neighbors of a node could be unacceptably long.

We now introduce the new version of the problem, called {\em Partial Cover Time with Budget:} any node in the graph is allowed to query only a fixed number of neighbors to retrieve their degrees.

For the budget version of the problem we now define the  \texttt{MD} algorithm.
\texttt{MD} algorithm combines ideas from the literature, i.e., it builds random walks that have the following properties: {\em (a)} unvisited neighbors are preferred and {\em (b)} among unvisited neighbors lowest-degree nodes are preferred.

\subsection{The \texttt{MD} algorithm}
\label{sub:min-degree}

We now describe our \texttt{MD} algorithm.
It takes as input an undirected and connected graph $\mathcal{G} = \langle N, E \rangle$ with $\vert N \vert = n$ nodes and $\vert E \vert = m$ edges, a threshold $\tau \in [0, 1]$, a starting node $i \in N$, and an integer budget $B$ whose meaning will be clarified later.
It returns the number of steps a random walk starting from $i$ needs to visit, at least once; a subset of nodes of $\mathcal{G}$ consisting of $n_{\max} = \lfloor \tau \times \vert N\vert \rfloor$ nodes (see Algorithm \ref{alg:MD} for a high level description). 

\begin{algorithm}
  \caption{The \texttt{MD} algorithm}
  
\begin{algorithmic}
\STATE $V \gets \{i\}$\;
\STATE $x_i \gets 1$\;
\STATE $n_{\max} \gets \lfloor \tau \times \vert N \vert\rfloor$\;
\STATE $k \gets i$\;

\WHILE{$\vert V \vert < n_{\max}$}
\STATE $L(k) \gets V - N(k)$\;
\IF {$\vert L_k \vert == 0$}
        \STATE Draw a node $j$ uniformly at random from $N(k)$\;
        \STATE $k \gets j$
\ELSE
        \IF {$\vert L_k \vert \geq B$}
                \STATE Draw a random sample $\hat{L}(k)$ of size $B$ from $L_k$\;
                \STATE Let $j$ be the smallest degree node in $\hat{L}(k)$\;
                \STATE $k \gets j$\;
        \ELSE
            \STATE Let $j$ be the smallest degree node in $L(k)$\;
            \STATE $k \gets j$\;
        \ENDIF
\ENDIF
\STATE Add $k$ to $V$\;
\STATE $x_i \gets x_i +1$\;
\ENDWHILE
\RETURN $x_i$\;
\end{algorithmic}
\label{alg:MD}
\end{algorithm}

\texttt{MD} uses an auxiliary variable $x_i$ (which is set equal to $1$ at the beginning) to record its progress. 
Let also $k$ be an auxiliary variable storing the currently-visited nod (initially, of course, $k=i$). 
In addition, \texttt{MD} uses a set $V$ to record the set of nodes already visited which, at the beginning, stores only the node $i$.

The \texttt{MD} algorithm is iterative and, at each iteration, it aims at adding a node to the set of visited nodes $V$; the algorithm stops as soon as the set $V$ reaches cardinality $n_{\max} = \lfloor \tau \times \vert N \vert\rfloor$.

We thus describe the operations carried out within each iteration. 
Variable $k$ contains the current node the walker is on and let $N(k)$ contain the set of neighbors of $k$. 

\texttt{MD} will checks whether there are nodes in $N(k)$ which have not yet been visited; to do so it builds the set $L(k) = V - N(k)$.

If the cardinality of $L(k)$ is zero, then, there are no unvisited nodes in $N(k)$.
Hence we select a random neighbour, say $j$, as in a Standard Random Walk. 

In contrast, suppose that $\vert L(k)\vert > 0$, i.e., there is at least one of the neighbors of $k$ which have not yet been visited. 
In this case, the \texttt{MD} algorithm has two options: 

\begin{itemize}
    \item[a)] the set $L(k)$ contains at least $B$ elements: the algorithm draws, uniformly at random, a subset $\hat{L}(k)$ of size $B$. The algorithm choose the lowest degree node from $\hat{L}(k)$ as the next node to move. 
    
    \item[b)] The set $L(k)$ contains no more than $B$ elements: in this case, the algorithm chooses the lowest degree node in $L(k)$ as the next node to move to.
In both the two cases, let $j$ be the next node to visit. The algorithm \texttt{MD} renames the node $j$ into $k$, which is the current on which the random walk is positioned.
\end{itemize}

The \texttt{MD} algorithm updates the set $V$ by adding the node $k$ and it increments by one the variable $x_i$. 
As previously noted, the process above stops if the cardinality of the set $V$ reaches $n_{\max}$, $x_i$ is returned  as output.

As observed in Section \ref{sec:background}, the number of steps a random walk starting from a node $i$ takes to visit a fraction of nodes of $G$ is a random variable $X_i$ and, thus, the output of the \texttt{MD} algorithm is a realization, called $x_i$, of $X_i$. 
If we apply \texttt{MD} a large number of times, say $T$, we generate a sequence of observed values $x_i^{1}, \ldots,x_i^{T}$ and we take their average:

\begin{equation}
    \label{eqn:mean-cover}
    \rho(\tau) = \frac{1}{T}\sum_{\ell =1}^{T} x_i^{\ell} 
\end{equation}

By the Strong Law of Large Numbers \cite{ross2006first}, we obtain that $\rho(\tau)$ converges to the actual partial cover time $PCT(\tau, i)$ (see Definition \ref{def:partial-cover-time}).
In our experiments we found that $T = 10$ was sufficient to ensure convergence.

\subsection{The role of the budget $B$}
\label{sub:budget}

The budget $B$ has a fundamental role in the \texttt{MD} algorithm that we wish to clarify in this section. 
When $B$ is set to 1 the algorithm chooses, uniformly at random, one of the unvisited neighbors of the current node and, thus, it coincides with the Edge Process algorithm described in \cite{Berenbrink*10}.

It is instructive to consider the behaviour of the \texttt{MD} algorithm as $B$ increases and, in detail, we wish to observe that if $B$ is sufficiently large, then the \texttt{MD} algorithm would degenerate into a deterministic procedure. 
Specifically, let us suppose that the \texttt{MD} algorithm is currently visiting the node $i$; for a fixed value of $B$, say $B = B^{\star}$, let $L^{\star}_i$ be the set of nodes from which \texttt{MD} will choose the next node to move. 

By construction, the \texttt{MD} algorithm selects the smallest degree node $n_{\min}^{\star} \in L^{\star}_i$. 
We ask for the probability $p$ that $n_{\min}^{\star}$ coincide with the smallest degree node $n_{\min}$ in $L_i$.

The estimation of $p$ depends on the node degree distribution and it will be experimentally discussed in Section \ref{sub:budget-tuning-experiment}; however, we expect that $p$ will increase if the ratio
$\frac{B^{\star}}{\vert L_i \vert}$ increases too.
At the limit case $B^{\star} = \vert L_i \vert$ such a probability should be equal to one. 
Therefore, if $B^{\star}$ approaches to $\vert L_i \vert$, then  \texttt{MD}  would {\em always direct} the walk to a pre-specified node (namely the unvisited node of lowest degree) and, thus, it could be no longer considered a proper random process.

%% file: experiments.tex
\section{Experimental Analysis}
\label{sec:experiments}

We have experimentally validated our \texttt{MD} algorithm by a comparative benchmark over a diversified set of six real datasets that are available in the public domain. 
We sought to address the following fundamental questions:

\begin{itemize}
    \item[$\mathbf{RQ_1}$] What is the optimal value for the budget $B$?
    
     \item[$\mathbf{RQ_2}$] How efficient is the \texttt{MD} algorithm to find a partial cover of a graph $G$ against other, state-of-the art, methods?
\end{itemize}

\subsection{Dataset Description}
\label{sub:dataset-description}
We used six publicly-available benchmark graphs, whose main features are summarized in Table \ref{tab:dataset-main-features}.

\textsc{\textbf{Facebook-Pages}}. 
This dataset was collected through the Facebook Graph API in November 2017 \cite{rozemberczki2019multi}. 
Nodes identify Facebook pages belonging to one of the following categories: politicians, governmental organizations, television shows and companies. 
Edges identify mutual ``likes'' between pages.

\textsc{\textbf{GitHub}}. 
This dataset was collected from the public GitHub API in June 2019 \cite{rozemberczki2019multi} and it  describes a social network of GitHub developers. 
Nodes are developers who have starred at least 10 repositories and edges identify mutual follower relationships between them.  

\textsc{\textbf{BrightKite}}.
This dataset was obtained by collecting all the public  check-in  data  between  April 2008  to  October   2010  for  BrightKite, a location-based social networking Web site \cite{ChoML11}. 
Nodes are associated with BrighKite members and edges specify friendship relationships. 

\textsc{\textbf{Facebook Friendship}}.
This dataset contains friendship data of Facebook users \cite{viswanath2009evolution}. 
A node represents a user and an edge represents a friendship between two users.

\textsc{\textbf{Flickr}}.
This dataset defines a graph in which nodes correspond to images from Flickr \cite{McAuley12}. 
Edges are established between images which share some metadata, such as the same location or common tags to annotate an image.

\textsc{\textbf{Amazon}}. 
This dataset defines the  Amazon  product co-purchasing network described in \cite{yang2015defining}.  
Nodes represent  products  and  edges  connect  commonly  co-purchased  products.

\begin{table*}
  \caption{Main features of the graphs used in our experimental evaluation}
  \label{tab:dataset-main-features}

  \begin{tabular}{ccccc}
    \toprule
    Dataset & Number of & Number of  &  Clustering & Diameter\\
    & Nodes & Edges & Coefficient & \\
    \midrule
    \textsc{\textbf{Facebook-Pages}} & 22\, 470 & 171\, 002 & 0.232 & 15\\
    \textsc{\textbf{GitHub}} & 37\, 700 & 289\, 003 & 0.013 & 7\\
\textsc{\textbf{BrightKite}} & 58\, 228 & 214\, 078 & 0.172 & 16\\  
     \textsc{\textbf{Facebook Friendship}} & 63\, 731 & 817\, 035 & 0.148 & 15\\      \textsc{\textbf{Flickr}} & 105\, 938 & 2\,316\, 948 & 0.089 & 9\\
    \textsc{\textbf{Amazon}} & 334\, 863 & 925\, 872 & 0.397 & 44\\
   
    \bottomrule
  \end{tabular}
\end{table*}

\begin{figure*}[htb]
  \centering  
  \begin{subfigure}{0.31\textwidth}
    \includegraphics [width=\textwidth] {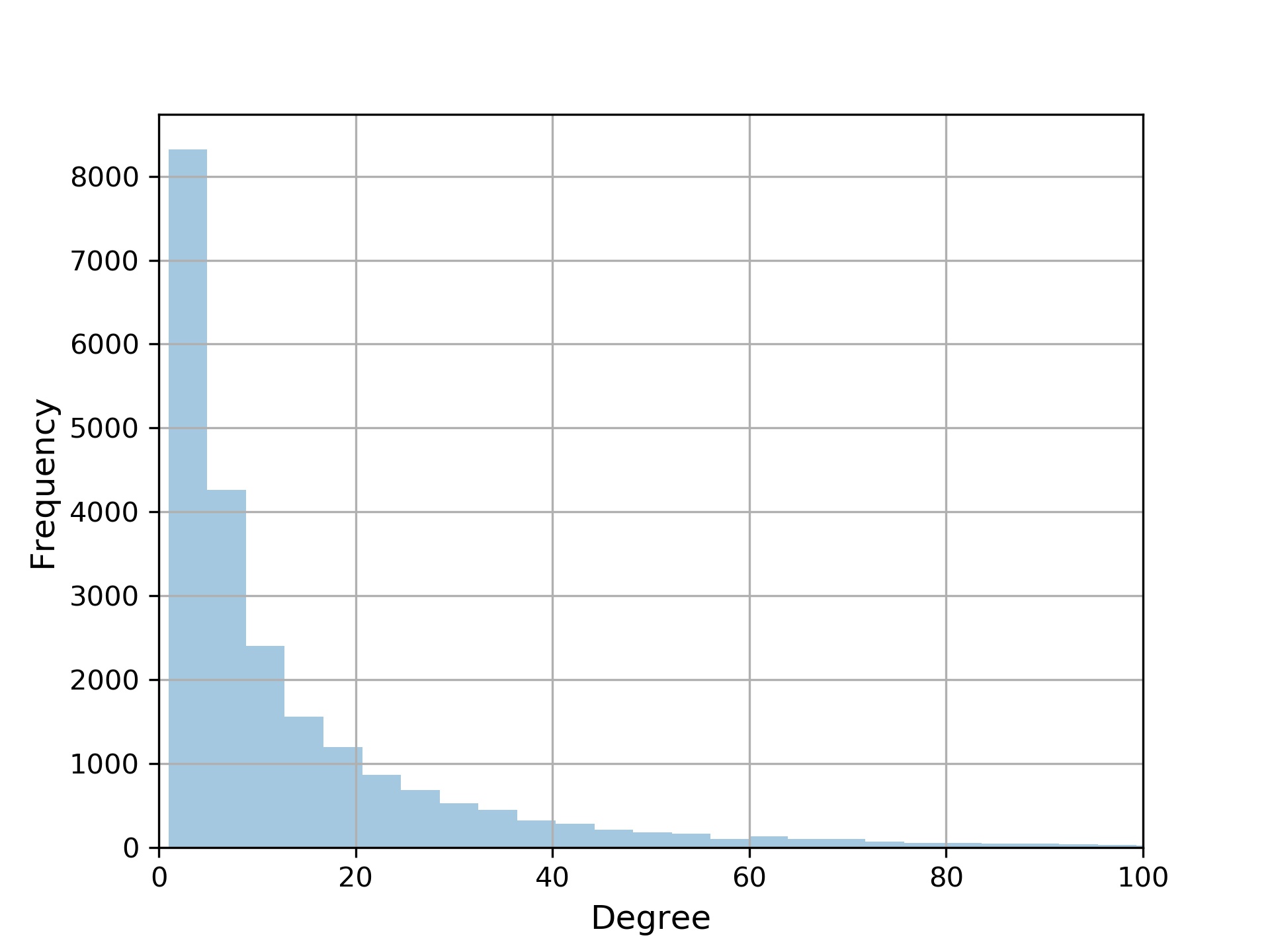}
    \vspace{-2em}
    \caption{\textsc{\textbf{Facebook Pages}}}
    \label{fig:FB-pages-degree}
  \end{subfigure}
  \begin{subfigure}{0.31\textwidth}
    \includegraphics [width=\textwidth] {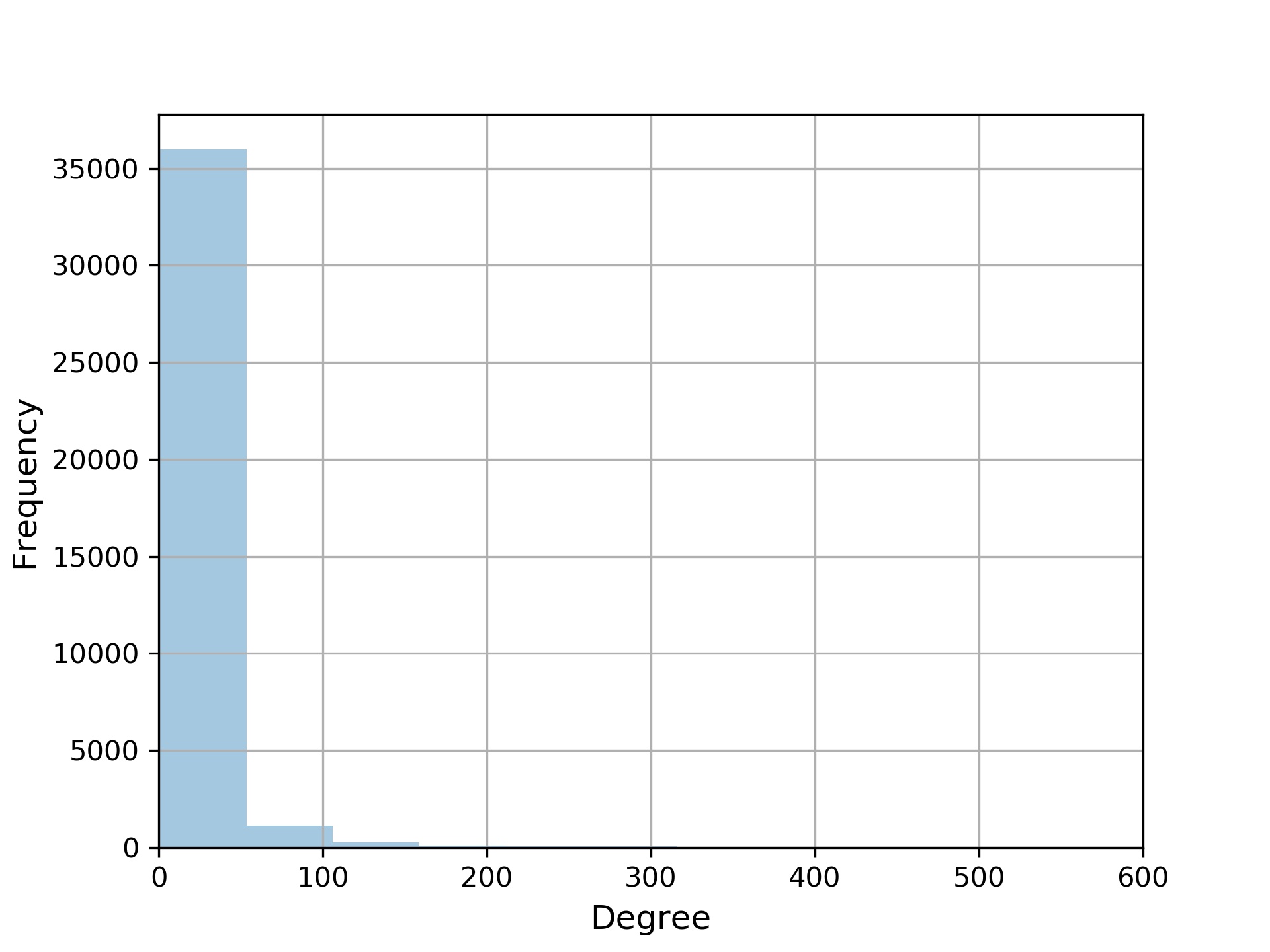}
    \vspace{-2em}
    \caption{\textsc{\textbf{GitHub}}}
    \label{fig:Git-degree}
  \end{subfigure}
  \begin{subfigure}{0.31\textwidth}
    \includegraphics [width=\textwidth] {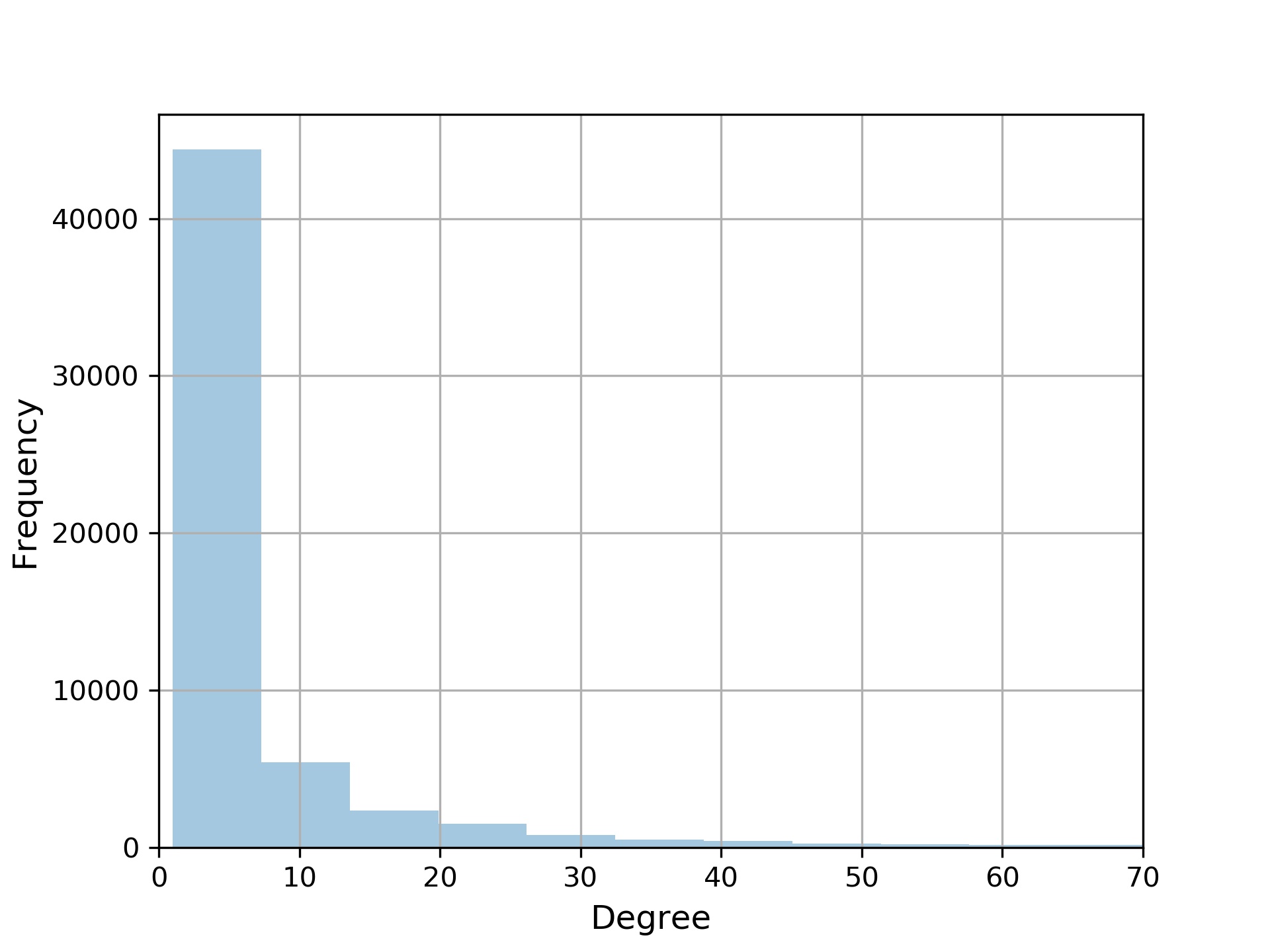}
    \vspace{-2em}
    \caption{\textsc{\textbf{BrightKite}}}
    \label{fig:Brightkite-degree}
  \end{subfigure}
  \begin{subfigure}{0.31\textwidth}
    \includegraphics [width=\textwidth] {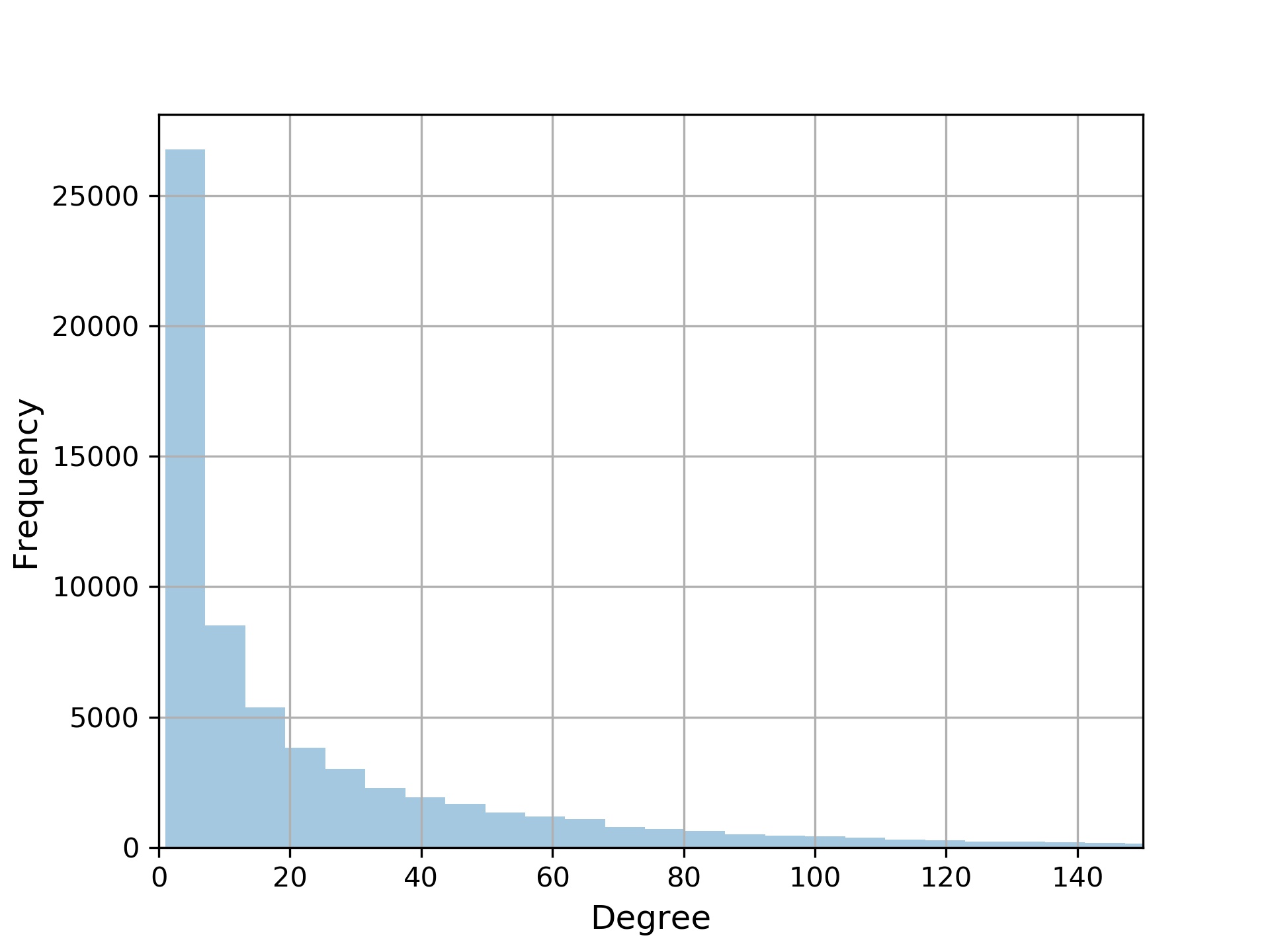}
    \vspace{-2em}
    \caption{Facebook}
    \label{fig:FB-links-degree}
  \end{subfigure}
  \begin{subfigure}{0.31\textwidth}
    \includegraphics [width=\textwidth] {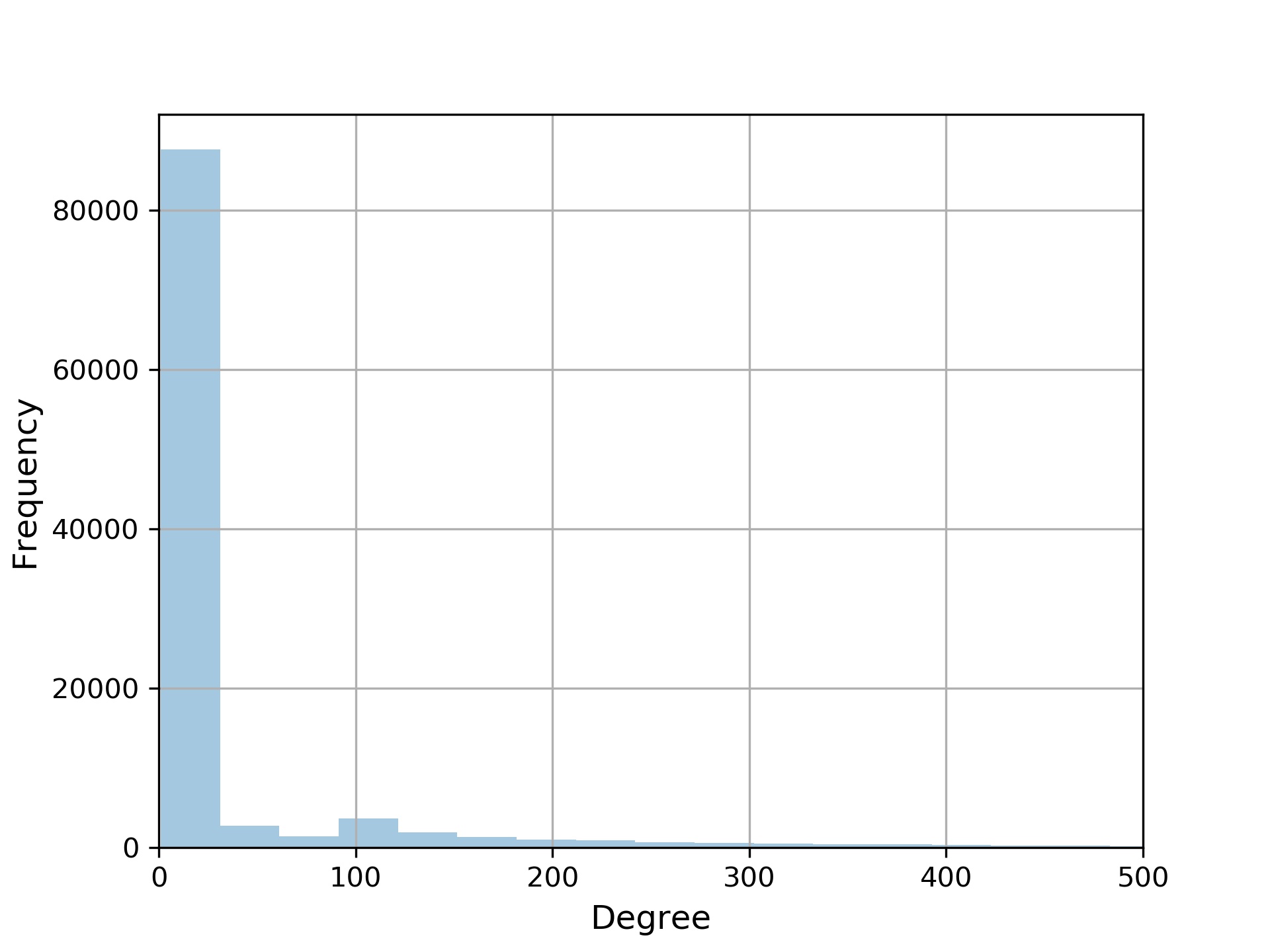}
    \vspace{-2em}
    \caption{\textsc{\textbf{Flickr}}}
    \label{fig:Flickr-degree}
  \end{subfigure}
  \begin{subfigure}{0.31\textwidth}
    \includegraphics [width=\textwidth] {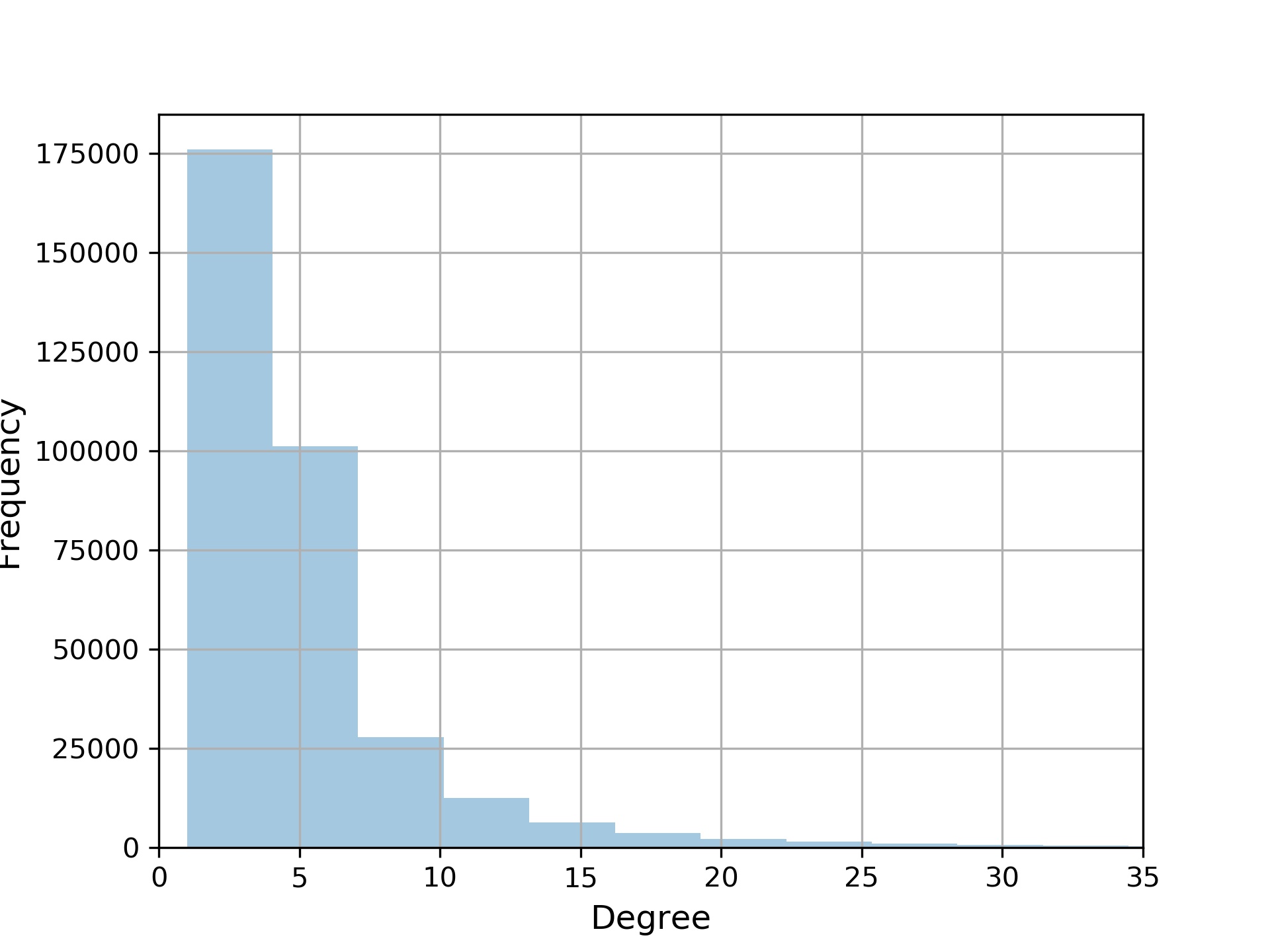}
    \vspace{-2em}
    \caption{\textsc{\textbf{Amazon}}}
    \label{fig:Amazon-degree}
  \end{subfigure}
  \vspace{-0.8em}
     \caption{Degree Distribution for the six datasets}
\label{fig:node-degree-distribution}
\end{figure*}

In Figures \ref{fig:node-degree-distribution}(\subref{fig:FB-pages-degree})- \ref{fig:node-degree-distribution}(\subref{fig:Amazon-degree}) we report node degree distribution for the datasets used in our tests.

We observe that node degree distribution is right-skewed for all datasets considered in our experimental trials.

Differences in observed distributions are likely to derive from the mechanisms regulating the formation and growth of each social network.
For instance, the \textsc{\textbf{GitHub}} dataset collects mutual likes between GitHub members who are quite active as software contributors, and, thus, the average node degree is higher than in other social networks and approximately thousand nodes have a degree ranging from 50 to 110. 
Other datasets such as \textsc{\textbf{Amazon}} have edges that represent the so-called \emph{co-purchase} relationship.
As expected, we observe that more than half of the nodes in the \textsc{\textbf{Amazon}} dataset display a degree less than five and the probability of observing a node with degree bigger than fifty is close to zero.  

\subsection{Evaluation Metrics}
\label{sub:evaluation}

We introduce the {\em normalized partial cover time $C(\tau)$ of an algorithm} as:

\begin{equation}
    \label{eqn:efficiency}
    C(\tau) = \frac{\rho(\tau)}{n}
\end{equation}

Here, the parameter $\rho(\tau)$ has been introduced in Equation \ref{eqn:mean-cover} and it is normalized by the number $n$ of graph nodes in order to make comparisons across graphs of different order possible.

Of course, the normalized partial cover time $C(\tau)$ increases (or, at least, it does not decrease) if $\tau$ increases. 
Given two methods $M_1$ and $M_2$ and a threshold $\overline{\tau} \in [0, 1]$, we say that the algorithm $M_1$ is {\em more efficient} than $M_2$ if the normalized partial cover time $C_1(\overline{\tau})$ associated with $M_1$ is less than the normalized partial cover time $C_2(\overline{\tau})$ associated with $M_2$.  

\subsection{Baseline Methods}
\label{sub:baseline-methods}

We compared the \texttt{MD} algorithm with four baseline algorithms from the literature, namely:

\begin{itemize}
    \item \textbf{Standard Random Walk, SRW}. 
    This is the well-known random walk over an undirected and connected graph in which the walker selects the next node to move uniformly at random among its neighbors.
    
    \item \textbf{Edge-Process, EP}. 
    This is the method described in \cite{BeCoFr15}, and, unlike SRW, the random walk prefers unvisited edges to select the next node to reach.
    
    \item \textbf{All Degrees, AD}. 
    This is the method described in \cite{ikeda2009hitting} and it assumes that a node knows the degree of all its neighbors. We recall that the AD method
    is {\em optimal for cover time}, i.e., it achieves a cover time of $O\left(n^2 \log n\right)$ independently of the topology of the graph.
    
    \item \textbf{Random Walks with Choice - RWC(d)}. 
    This is the method described in \cite{avin2008power}; in compliance with recommendations provided in \cite{avin2008power} and after some experiments, we decided to set $d = 3$ because such a value of $d$ offered the lowest $C(\tau)$.
\end{itemize}

All these methods have been described in Section \ref{sec:related-works}. 
We also tried the method described in \cite{abdullah2015speeding} but we found it had worse performance than other methods above and, thus, we do not report its results here.

\subsection{Budget Tuning ($RQ_1$)}
\label{sub:budget-tuning-experiment}

In this section we study the role of the budget $B$ on our \texttt{MD} algorithm. 
Recall from Section \ref{sub:budget} when $B$ increases the probability $p$ that will \texttt{MD} choose the smallest-degree node among the neighbors will increase accordingly; so for higher values of B  \texttt{MD} could be no longer considered a random-search process. 

\begin{figure}[htbp]
    \centering
    \includegraphics[width=0.47\textwidth]{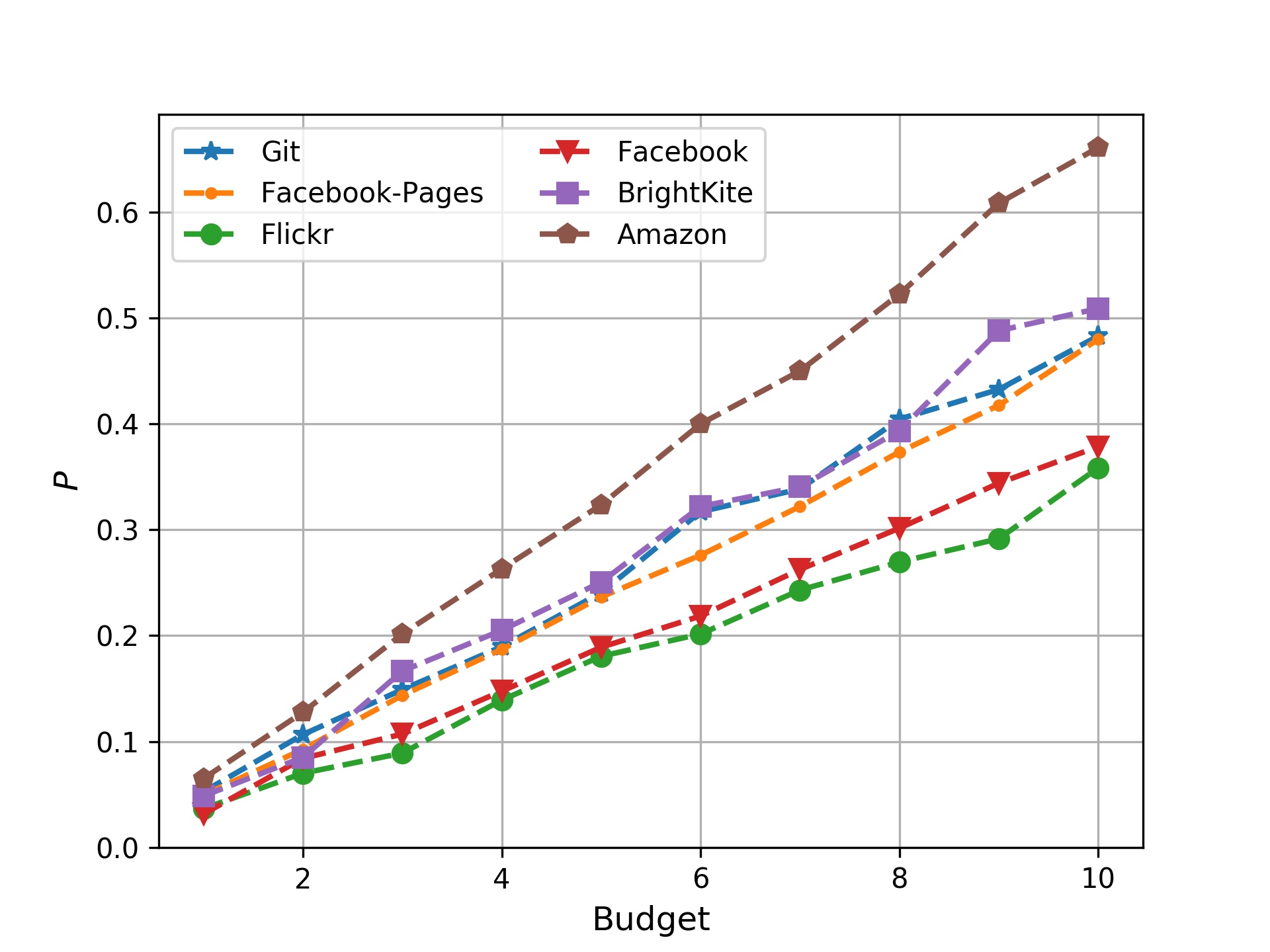}
    \caption{Probability $p$ of selecting the smallest degree node as function of the budget $B$.}
    \label{fig:probability}
\end{figure}

Figure \ref{fig:probability} reports the values of $p$ as function of the budget $B$ for all our datasets. 
Firstly, observe that, for all datasets under scrutiny, a value of $B = 10$ corresponds to a probability $p$ ranging from 0.37 to 0.66: in other words, the \texttt{MD} algorithm has a high chance of discovering (and, thus, selecting) the smallest degree node even if it has at its disposal only ten nodes. 
Such a behavior depends on the degree distribution observable in many real-life graphs and, in particular, in the graphs considered in our study: node degree distribution is, in fact, right-skewed which implies that the vast majority of nodes displays a small degree (generally less than five). 
Therefore, a random sample of nodes in one of our graphs will, with high probability, contain one or more nodes of small degree; in many cases, the sample will also contain a node showcasing minimum degree.

A further observation is that $p$  grows linearly with $B$ in all datasets but its \emph{rate of growth} differs across  datasets: the steepest increase in $p$ occurs for  \textsc{\textbf{Amazon}}. 
Differences in slopes are attributable to the different node degree distribution we observe in each graph.

Finally, Figure \ref{fig:probability} suggests that a value of $B = 5$ is generally reasonable because it implies a value of $p$ always less than 0.32. 
Therefore, we set $B = 5$ for the experiments next.

\subsection{Performance comparison ($RQ_2$)}
\label{sub:performance-comparison}

We used the normalized partial cover time $C(\tau)$ to compare the methods introduced in Section \ref{sub:baseline-methods} and the \texttt{MD} algorithm.

\begin{figure*}[htb]
  \centering  
  \begin{subfigure}{0.47\textwidth}
    \includegraphics [width=\textwidth] {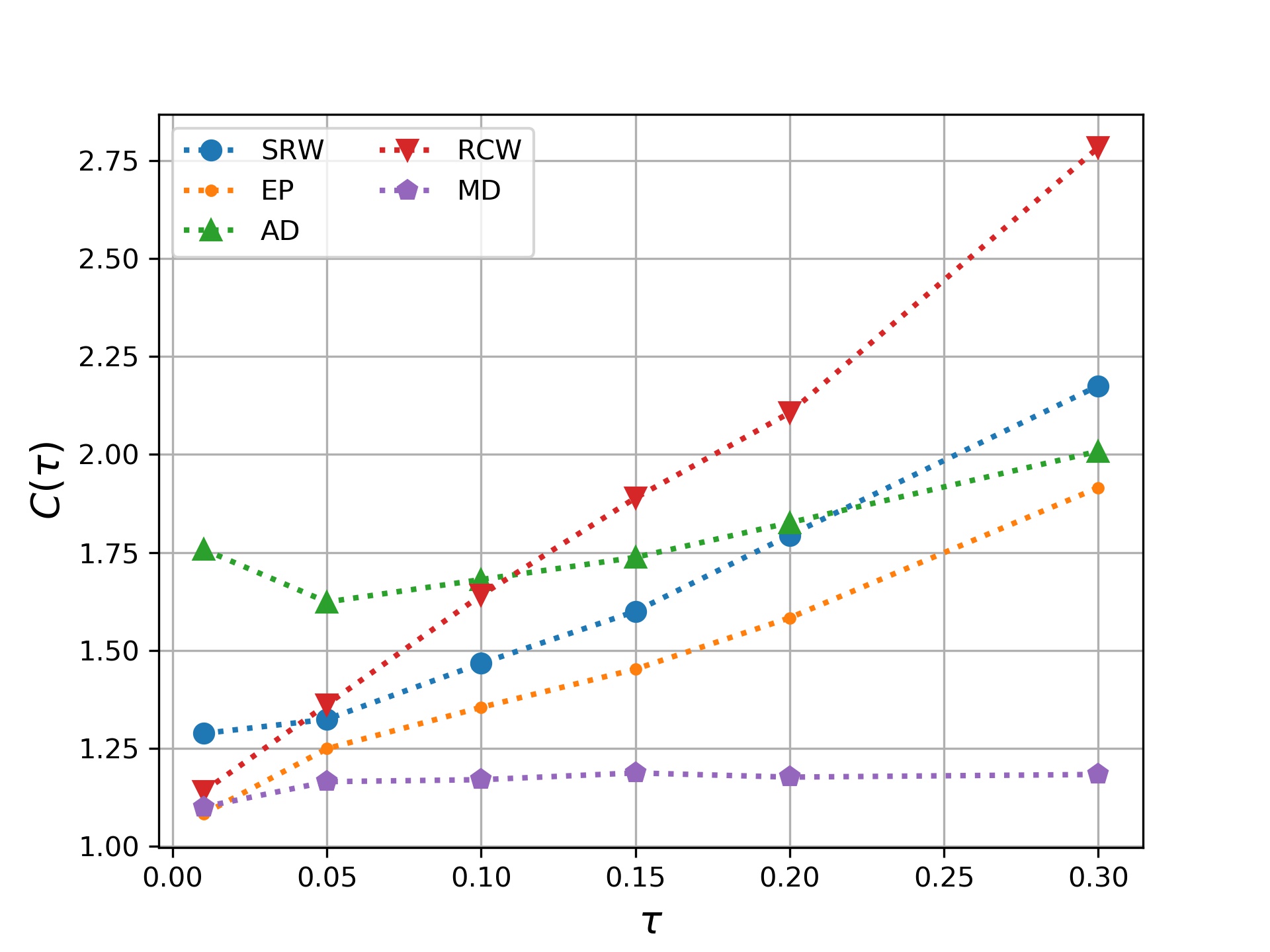}
    \vspace{-2em}
    \caption{\textsc{\textbf{Facebook Pages}}}
    \label{fig:FB-pages}
  \end{subfigure}
  \begin{subfigure}{0.47\textwidth}
    \includegraphics [width=\textwidth] {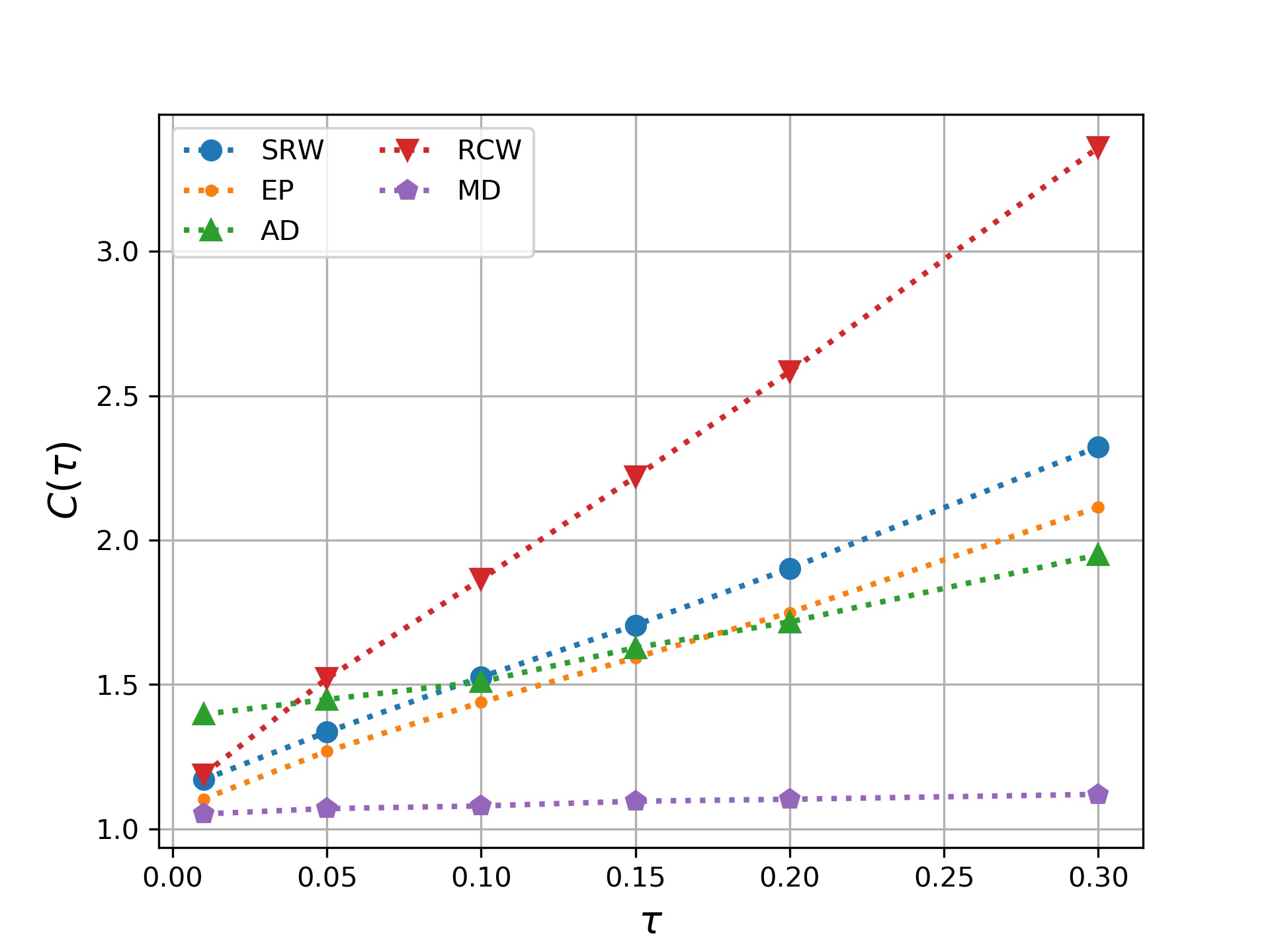}
    \vspace{-2em}
    \caption{\textsc{\textbf{GitHub}}}
    \label{fig:Git}
  \end{subfigure}
  \begin{subfigure}{0.47\textwidth}
    \includegraphics [width=\textwidth] {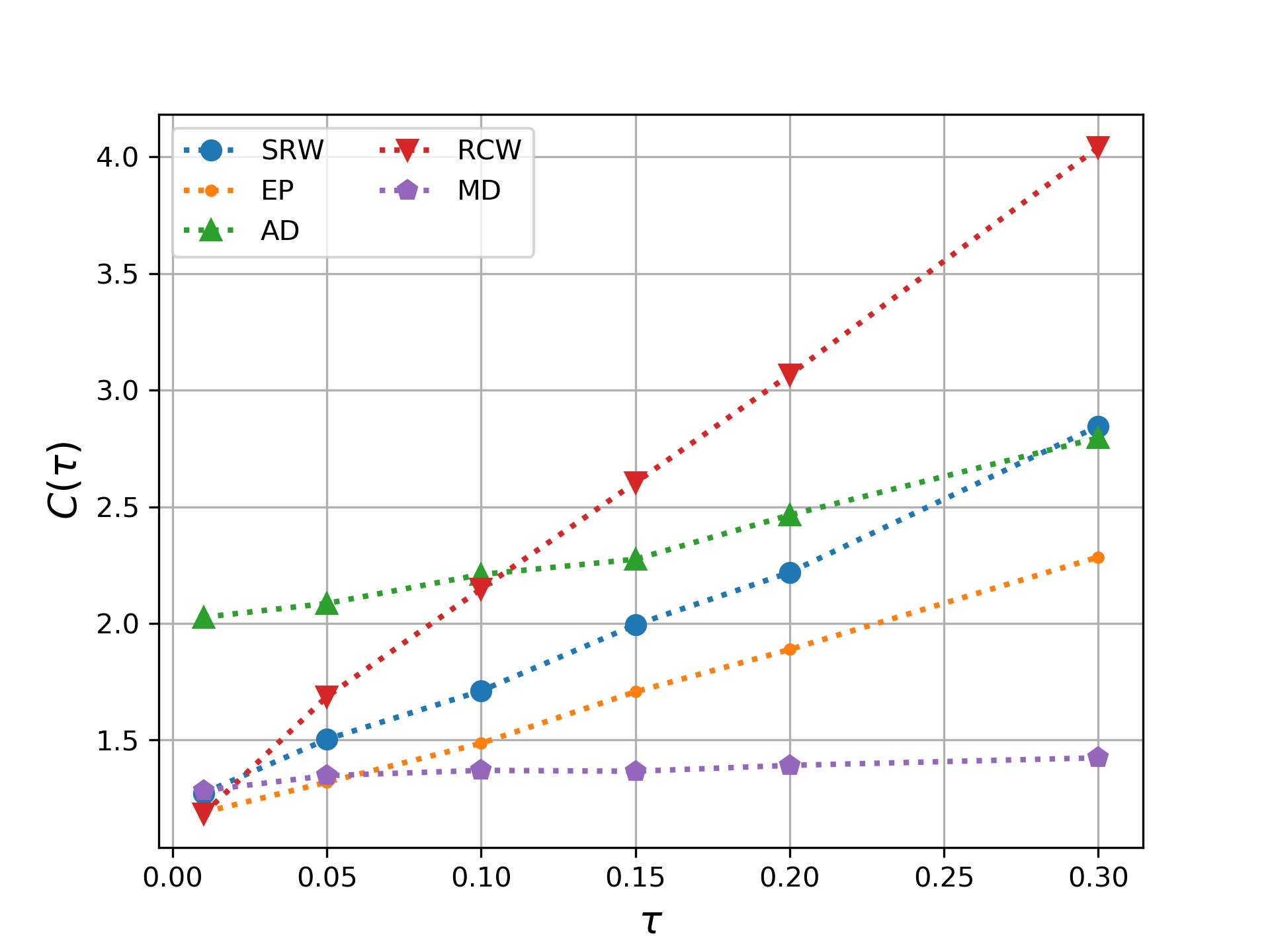}
    \vspace{-2em}
    \caption{\textsc{\textbf{BrightKite}}}
    \label{fig:Brightkite}
  \end{subfigure}
  \begin{subfigure}{0.47\textwidth}
    \includegraphics [width=\textwidth] {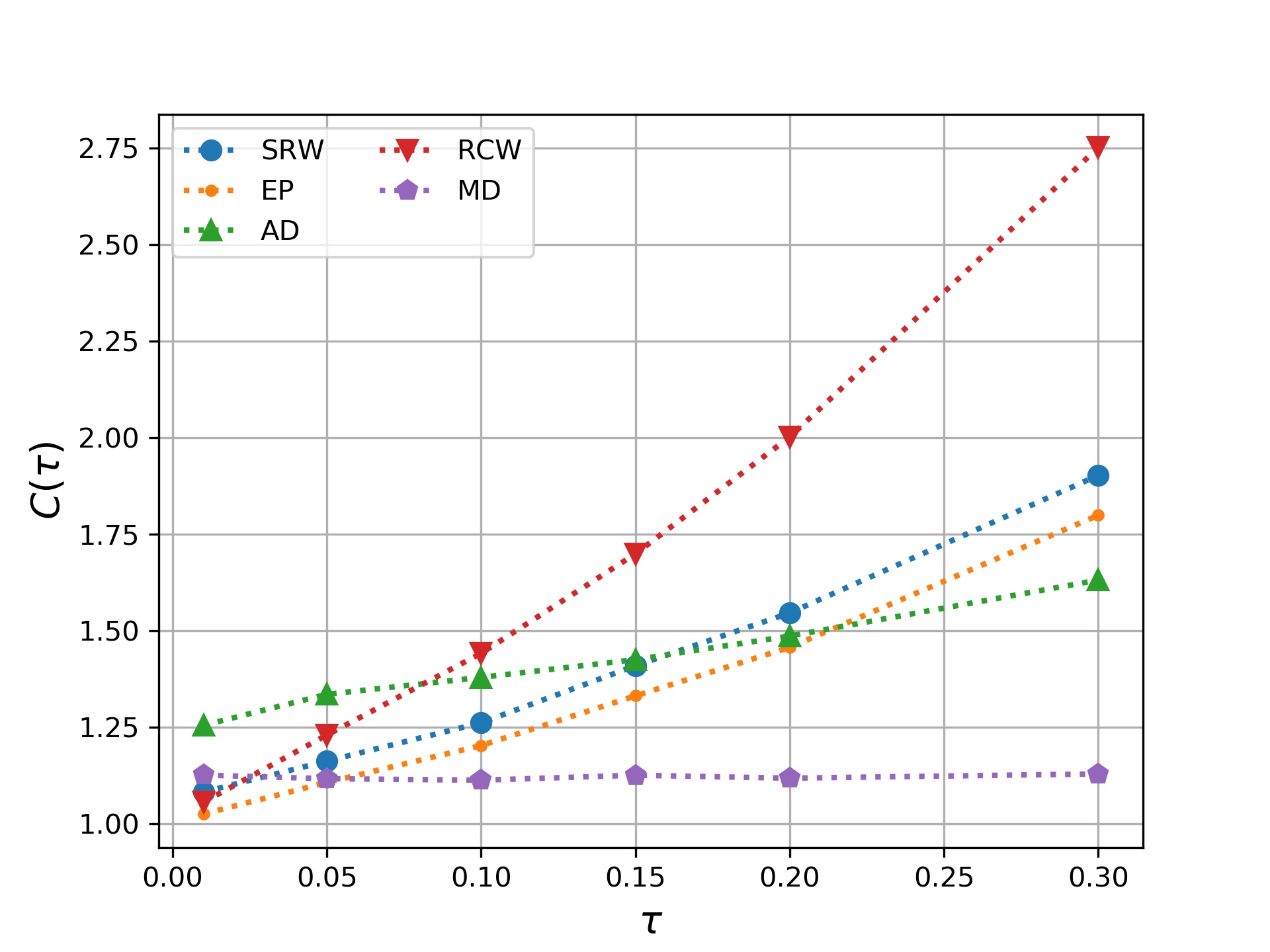}
    \vspace{-2em}
    \caption{Facebook}
    \label{fig:FB-links}
  \end{subfigure}
  \begin{subfigure}{0.47\textwidth}
    \includegraphics [width=\textwidth] {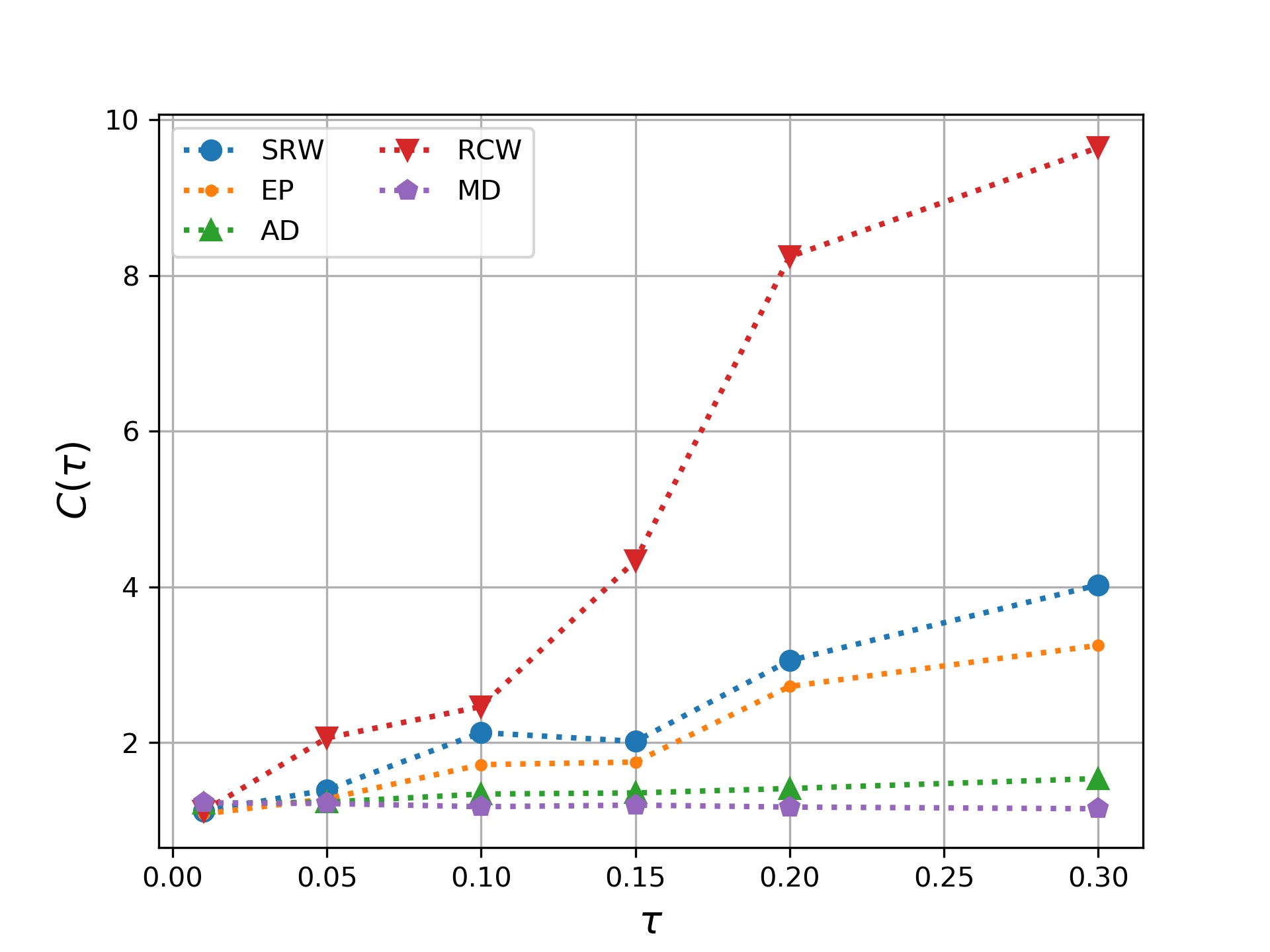}
    \vspace{-2em}
    \caption{\textsc{\textbf{Flickr}}}
    \label{fig:Flickr}
  \end{subfigure}
  \begin{subfigure}{0.47\textwidth}
    \includegraphics [width=\textwidth] {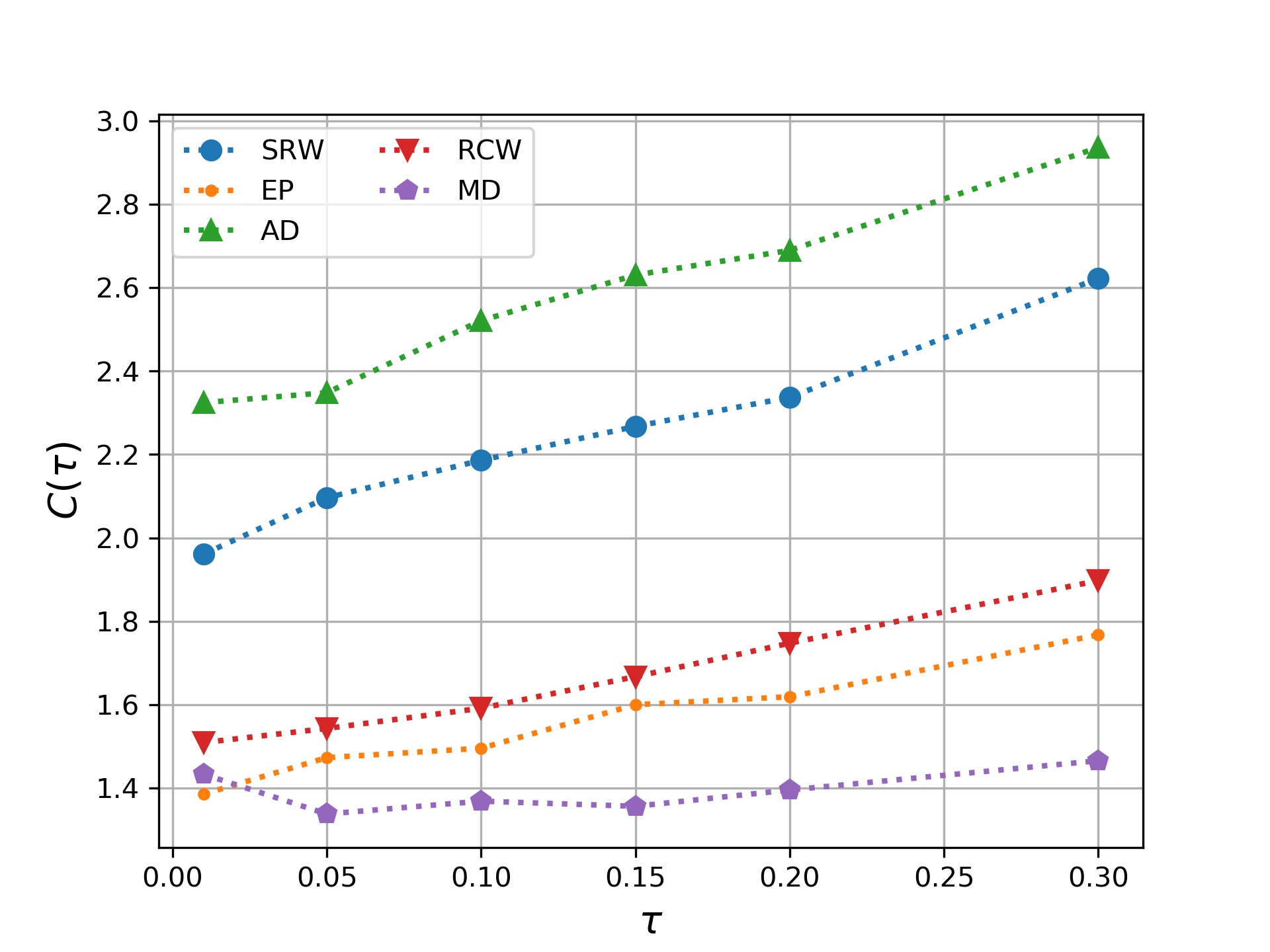}
    \vspace{-2em}
    \caption{\textsc{\textbf{Amazon}}}
    \label{fig:Amazon}
  \end{subfigure}
  \vspace{-0.6em}
     \caption{Values of the normalized partial cover time $C(\tau)$ as function of $\tau$ for the SRW, EP, AD, RCW and \texttt{MD} algorithms.}
\label{fig:efficiency-vs-tau}
\end{figure*}

The normalized partial cover time $C(\tau)$ obtained for values of $\tau$ ranging from 0.01 to 0.3 is reported in Figures \ref{fig:efficiency-vs-tau}(\subref{fig:FB-pages})- \ref{fig:efficiency-vs-tau}(\subref{fig:Amazon}). 
The main findings of our experimental analysis can be summarized as follows:

\begin{itemize}
    \item The \texttt{MD} algorithm significantly outperforms all other approaches if $\tau > 0.05$. 
    In contrast, if $\tau \leq 0.05$ and we concentrate on \textsc{\textbf{Brightkite}}, \textsc{\textbf{Facebook Friendship}} and \textsc{\textbf{Amazon}} datasets, the \texttt{MD} algorithm  is suboptimal, even if its normalized partial cover time $C(\tau)$ is very close to that of the best performing methods. 
    
    The increase of $C(\tau)$ due to the increase $\tau$ in the \texttt{MD} algorithm is generally much slower than that experienced by other methods. 
    We can therefore confirm the algorithmic idea underpinning \texttt{MD}, i.e., that biasing random walks toward low-degree and unvisited nodes actually accelerates the process of visiting a graph.
    
    \item Apart from our approach, the EP method performs very well if $\tau$ is small (i.e., it is smaller than $0.1$); if $\tau$ is larger than $0.1$ and we focus on the \textsc{\textbf{Facebook}} and \textsc{\textbf{Flickr}} datasets, the normalized partial cover time associated with the EP method deteriorates significantly but it is often significantly better than the normalized partial cover time observed for other methods. 
    We can conclude, therefore, that the strategy of privileging unvisited edges yields a remarkable acceleration.
    
    \item In the SRW approach, we report an almost linear increase in $C(\tau)$ as $\tau$ increases too. 
    If $\tau$ is smaller than $0.05$, the SRW method is competitive with other methods, with the exception of the \textsc{\textbf{Amazon}} dataset. 
    In general, poor performances of the SRW algorithm depends on the fact that the algorithm visits a node more than once and, thus, a larger number of steps are required before terminating.
    
    \item The AD method performs very well on the \textsc{\textbf{Flickr}} dataset: here, its normalized partial cover time is close to that of the \texttt{MD} algorithm and it is significantly smaller than the normalized partial cover time of all other methods. \textsc{\textbf{Flickr}} is also the most arduous dataset among those under scrutiny, i.e., the dataset on which all methods under investigation showcase the worst values of the normalized partial cover time. 
    The AD method displays its worst performances on the \textsc{\textbf{Amazon}} dataset.
   That's not surprising: while the AD algorithm achieves, in the worst case, the optimal cover time for a graph of arbitrary topology, there are no guarantees that AD will is also be the most efficient choice for minimizing the partial cover time (and, thus, for normalized partial cover time) \cite{avin2004efficient}. 
   Our experiments, therefore, prove that on real-life graphs the AD algorithm might not be competitive, if we goal is to minimize the (unbudgeted) normalized partial cover.
    
    \item With the exception of the \textsc{\textbf{Amazon}} dataset, the normalized partial cover time of the $RCW(d)$ algorithm is worse than that of all other methods. 
    This result is somewhat surprising since the normalized partial cover time of the $RCW(d)$ algorithm is often worse than that of an SRW. 
    
    It must be stressed, however, that the $RCW(d)$ approach has been designed to optimize the partial cover time for specific topologies such as regular graphs, grids, hypercubes or random geometric graphs (used to model wireless networks). 
    Those topologies differ significantly from the topology of the graphs considered in our study (which display a high irregularity in the node degree distribution). 
    Differences in graph topology have a big impact on the partial cover time and they explain the large values of $C(\tau)$ we observed for the $RCW(d)$ algorithm.    
\end{itemize}

%% file: conclusions.tex
\section{Conclusions}
\label{sec:conclusions}
We have introduced a variation of the (Partial) Graph Cover Time problem that considers budgets, defined as a limit on the accessibility of neighbor nodes. 
We have designed an efficient random-walk solution which operates exactly under the constraint that a node can access only a fraction of its neighbors.
The \texttt{MD} algorithm introduce here favorably combines heuristic search ideas, namely the preference for unvisited nodes and, among those, for lowest-degree ones.
Experiments on six real-life graphs in the Social Web scenario have confirmed that \texttt{MD} is effective and can outperform state-of-the-art methods.

We plan to extend these results in several directions. 
For instance, we will assess how (and if) partial cover time $PCT_G(\tau, i)$ depends on the choice of the ``start'' node  and, specifically, whether some node features associated with a start node $v$ (such as the Betweenness Centrality or the Eigenvector Centrality of $i$) are in fact predictive of $PCT_G(\tau, v)$. 
Another direction of research is to parametrize the number of allowed visits to metadata on size and density of the input graph, e.g. with a log-log dependence between number of nodes and budget (which corresponds to considering \texttt{MD} optimal for $|\mathcal{G} | > 2^{16}\approx 64k$). 
Finally, we will propose an analogous of \texttt{MD} for the directed-graph case, which is the natural model for asymmetric relationships that we had previously explored such as trust networks \cite{Agreste*15} and online negotiation \cite{Costantini13}.